\documentclass[12pt,fleqn]{article}
\usepackage{amsmath,amssymb,amsthm,epsfig,graphics}
\usepackage{lscape}

\newcommand{\bab}{\end{gather}}
\newcommand{\ri}{{\mathrm i}}
\newcommand{\p}{\partial}
\newcommand{\bea}{\begin{array}}
\newcommand{\eea}{\end{array}}
\newcommand{\beg}{\begin{gather}}

\catcode`@=11 \long
\def\@caption#1[#2]#3{\par\addcontentsline{\csname
ext@#1\endcsname}{#1} {\protect\numberline{\csname
the#1\endcsname}{\ignorespaces #2}} \begingroup \small
\@parboxrestore \@makecaption{\csname fnum@#1\endcsname}
{\ignorespaces #3}\par \endgroup} \catcode`@=12

\newcommand{\C}{\mathbb{C}}

\newcommand{\la}{\label}

\catcode`@=11 \long
\def\@caption#1[#2]#3{\par\addcontentsline{\csname
ext@#1\endcsname}{#1} {\protect\numberline{\csname
the#1\endcsname}{\ignorespaces #2}} \begingroup \small
\@parboxrestore \@makecaption{\csname fnum@#1\endcsname}
{\ignorespaces #3}\par \endgroup} \catcode`@=12

\begin{document}

\allowdisplaybreaks
 \begin{titlepage} \vskip 2cm

\begin{center} {\Large\bf Superintegrable  systems with position dependent mass}

 \vskip 3cm {\bf {A. G. Nikitin }\footnote{E-mail:
{\tt nikitin@imath.kiev.ua} }
\vskip 5pt {\sl Institute of Mathematics, National Academy of
Sciences of Ukraine,\\ 3 Tereshchenkivs'ka Street, Kyiv-4, Ukraine,
01601\\}}
\vskip 1cm{\bf T. M. Zasadko}
\footnote{E-mail:{\tt tacita@ukr.net}} \vskip 5pt{\sl Taras Shevchenko National University of Ukraine,\\64 Volodymirska Street, Kyiv-4, Ukraine\\}\end{center}
\vskip .5cm \rm
\begin{abstract} {First order integrals of motion for Schr\"odinger  equations with position dependent masses are classified. Eighteen classes of such equations with non-equivalent symmetries are specified. They include integrable, superintegrable and maximally superintegrable systems. Among them is a system invariant with respect to the Lie algebra of Lorentz group and a system whose integrals of motion form algebra so(4). Three of the obtained systems are solved exactly.}\end{abstract}
\end{titlepage}

\allowdisplaybreaks

\section{Introduction\label{int}}Schr\"odinger equations are  perfect subjects for demonstration of the richness and profundity of the concept of symmetry. In addition to symmetries  with respect to continuous groups, they admit supersymmetries and  hidden symmetries like the Fock symmetry of the Hydrogen atom. Moreover, just Schr\"odinger equations belong to the nice field of the inverse problem approach with its infinite number of symmetries and constants of motion.

The history of searching for symmetries of Schr\"odinger equation is both long and inspiring. The invariance of this equation  w.r.t. Galilei group was in fact predicted by Sophus Lie. More exactly, Lie found the maximal continuous invariance group of the heat equation, which in the main coincides with the symmetry group of the free Schr\"odinger equation.

A systematic search for Lie symmetries of Schr\"odinger equation was carried out in papers \cite{Hag}, \cite{Nied}, \cite{And}, \cite{Boy} where the maximal invariance groups of this equation with arbitrary potential were presented. The next level symmetries, i.e., the second order symmetry operators for 2$d$ and 3$d$ Schr\"odinger equation have been found in \cite{wint1}, \cite{BM} and \cite{ev1}, \cite{ev2}. Symmetry operators of arbitrary order for the free Schr\"odinger equation had been enumerated in \cite{N7},  the completed group classification of the nonlinear Schr\"odinger equations was presented in \cite{pop}.

The extended (in particular, second order) symmetries are requested
for  description of systems admitting solutions in separated
variables \cite{Miller}, integrable and superintegrable systems \cite{wint01}. A relatively new field is presented by
superintegrable systems with spin whose systematic investigation
was started with paper \cite{w6} and continued in \cite{w7},
\cite{w8}, \cite{N2}, \cite{N3} and \cite{N4}, see also
survey \cite{wint01}. The subject of the research carried out
in \cite{w7}, \cite{w8} are systems with spin-orbit interaction
while the systems with Pauli type interactions are studied
in \cite{N2}, \cite{N3} and \cite{N4}.

Let us note that the first example of a superintegrable system with spin 1/2 was presented earlier in paper \cite{Pron1}. Superintegrable systems with arbitrary spin were discussed in \cite{Pron2}, \cite{N1}, \cite{N6} and \cite{N5}, the relativistic systems were elaborated in \cite{N2} and \cite{N77}.

 In the present paper we discuss superintegrability aspects of position dependent mass (PDM) Schr\"odinger equations which attract interest of many researchers and are applied in several physical systems. They are requested for description of  various condensed-matter systems such as semiconductors \cite{Roz}, \cite{1}, quantum liquids \cite{7} and metal clusters \cite{13}, quantum wells, wires and dots \cite{2}, \cite{3},  supper-lattice band structures \cite{5} and many, many others.

There are numerous papers devoted to exact solutions of PDM quantum mechanical problems, see, e.g., \cite{9}, \cite{11} \cite{10} and the references therein. The very possibility to solve exactly a mechanical or quantum mechanical equation is caused by its symmetry. And it is the reason why symmetries (and supersymmetries) of particular PDM problems were in the focus of attention of many researches, see, e.g., \cite{10}, \cite{raca}, \cite{Kocha}, \cite{Cruz}.  However, in contrast with the case of constant mass,  there are no general results concerning the group classification of generic PDM problems   and complete sets of their  integrals of motion. An exception is paper \cite{LL} where the generic form of PDM Schr\"odinger equation compatible with the Galilei invariance postulate is presented. In addition, starting with classical systems defined in  curved spaces, the extended  and  well defined class of PDM superintegrable models was found and studied in papers \cite{balt}, \cite{balt0}, \cite{balt2}, \cite{balt3} and \cite{balt4}, see also the references cited therein.

In the present paper we start a systematic investigation of integrable and superintegrable systems with position dependent mass. As the first step we classify PDM Schr\"odinger equations which admit the first order integrals of motion. Since these equations are defined up to two arbitrary functions, the level of complexity of the classification problem is compatible with the one characterizing  papers \cite{w6} and \cite{N3} where the first order integrals of motion for the Schr\"odinger equations with constant masses and matrix potentials  were classified.

It will be shown that there exist 18 non-equivalent classes of PDM Schr\"odinger equations with non-trivial integrals of motion. Among them there are superintegrable systems  invariant with respect to the Lie algebras of Lorentz group SO(1,3) and rotation group  SO(4). We present solutions of these systems which appears to be exactly solvable. Moreover, the spectra of the related Hamiltonians can be found algebraically like in the case of the Hydrogen atom.

\section{Determining equations}
Let us consider a stationary PDM Schr\"odinger equation
\begin{equation}\label{se}
   \hat H \psi=E \psi,
\end{equation}
where $H$ is  the following generic Hamiltonian
\begin{gather}\label{H}\hat H=p_af({\bf x})p_a-V({\bf x})=-\p_af({\bf x})\p_a -V({\bf x}).\end{gather}
Here ${\bf x}=(x^1,x^2,x^3),$ $p_a=-i\p_a$, $V({\bf x})$ and $f({\bf x})=\frac1{2m({\bf x})}$ are arbitrary functions associated with the effective potential and inverse effective PDM, and summation from 1 to 3 is imposed over the repeating index $a$.

Let us search for first order integrals of motion for equation (\ref{se}), i.e., for commuting with $H$ differential operators of first order which we write in the form:
\begin{equation}\label{Q}
    Q=\frac12(\xi^ap_a+p_a\xi^a)+\tilde\eta =-\ri(\xi_a\p_a+\eta)
\end{equation}
where $\eta=\frac12\p_a\xi^a+\ri\tilde\eta,$ $\xi^a$ and $\eta$ are functions  of $\bf x$.

By definition, operators $H$ and $Q$ should commute each other:
\begin{equation}\label{HQ}[\hat H,Q]\equiv \hat H Q-Q\hat H=0.\end{equation}

Since the commutator of  first order differential operators is again a first order differential operator, the complete set of integrals of motion (\ref{Q}) form a basis of a Lie algebra. Thus in fact we are searching for invariance algebras of equation (\ref{se}). Moreover, these algebras can be integrated to local Lie groups.

Relation (\ref{HQ}) has to be treated as an operator equation. It means that this differential expression should nullify any twice differentiable function. In other words, to solve this equation it is necessary to equate to zero the coefficients for all distinct differentials in (\ref{HQ}). In this way, after calculating the commutator, we obtain the following system of determining equations:
\begin{gather}
\xi^cf_{c}\delta_{ab}-f(\xi^b_{a}+\xi^a_{b})=0,\la{de0}
 \\\label{de3}
-\xi^if_{ai}+f_{i}\xi^a_{i}+f\xi^a_{cc}+2 f{\eta}_{a}=0, \\\label{de4}
f_{a}{\eta}_{a}+f{\eta}_{aa}-\xi^a{V}_{a}=0.
\end{gather}
Here $\delta_{ab}$ is the Kronecker symbol, and subindices denote derivations with respect to the corresponding spatial variable: $\eta_a=\p_a\eta$, etc.

The system of equations (\ref{de0})--(\ref{de4}) gives the necessary and sufficient conditions for commutativity of operators $H$ and $Q$. It is overdetemined and includes ten equations for six unknowns.

Let us evaluate the determining equations. First we note that equation (\ref{de0}) can be decoupled to two subsystems:
\begin{gather}\label{de1}\xi^b_{a}+\xi^a_{b}=\frac23\delta_{ab}\xi^i_i,\\
\label{de2}
3\xi^if_i=2f\xi^i_i\end{gather} which are nothing but the traceless part and  trace of (\ref{de0}).

Equation (\ref{de1}) defines the 3$d$ conformal Killing vector, whose general form is given by the following expression (see, e.g., \cite{N8})
\begin{gather}\label{sol1}
    \xi^a=\lambda^ax^nx^n-2x^a\lambda^nx^n+\mu^c\varepsilon^{abc}x^{b}+
    \omega x^a+\nu^a
\end{gather}
where the Greek letters denote arbitrary real constants.

Thus we already know the general form of functions $\xi^a$ which are second order polynomials in $x^a$. To find the remaining function $\eta$ needed to fix $Q$ it is sufficient to differentiate (\ref{de2}) w.r.t. $x_a$ and compare the result with (\ref{de3}).  In this way, using equations (\ref{de1}) and the following identities
\begin{gather*}\xi^i_{aa}=2\lambda_i,\quad \xi^i_{ia}=-6\lambda_a\end{gather*} we immediately find, that
\begin{gather}\la{sol2}\eta=-3\lambda_ax_a+c\end{gather} where $c$ is a constant.

If relations (\ref{sol1}),  (\ref{sol2}) and (\ref{de2}) hold then (\ref{de3}) and (\ref{de1}) are satisfied identically, while the remaining equations (\ref{de4}) and (\ref{de2}) are reduced to the following form:
\begin{gather}\la{de5}\xi^if_i=2(\omega-2\lambda_ax_a)f,\\\la{de6}
\xi^iV_i=-3\lambda_if_i\end{gather} where $\xi^i$ are functions  given in (\ref{sol1}). Thus our classification problem is reduced to finding the general solution of equations (\ref{de5}) and (\ref{de6}) for unknowns $f$ and $V$, where $\xi^i$ are polynomials (\ref{sol1}).
\section{Nonequivalent versions of equations (\ref{de5}) and (\ref{de6})}
Equations (\ref{de5}) and (\ref{de6}) include ten arbitrary parameters. Our next task is to specify values of these parameters which correspond to non-equivalent  versions of these equations.

To define equivalence relations for solutions of system (\ref{de5}), (\ref{de6}) we note that in accordance with (\ref{Q}), (\ref{sol1}), (\ref{sol2}) the generic first order integral of motion for equation (\ref{se}) can be represented as the following linear combination:
\begin{gather}\label{Q1} Q=\lambda^i K^{i}+\mu^iJ^i+\omega D+\nu^iP^i+c\end{gather}
where
\begin{gather}\label{QQ}\begin{split}&
 P^{i}=p^{i}=-i\frac{\partial}{\partial x_{i}},\quad J^{i}=\varepsilon^{ijk}x^jp^k, \\&
D=x^n p^n-\frac{3\ri}2,\quad K^{i}=x^nx^n p^i -2x^iD,\end{split}
\end{gather}
 and  $c$ is a constant  which is not essential and will be chosen as zero.

Operators (\ref{QQ}) satisfy the following commutation relations:
\begin{gather}\begin{split}\label{c(3)}&[P^a,P^b]=0, \quad [P^a,J^b]=\ri\varepsilon_{abc}P^c,\\&[J^a,J^b]=\ri\varepsilon_{abc}J^c,
\quad [D,J_a]=0,\\&[D,P^a]=\ri P^a,\quad [D,K^a]=-\ri K^a,\\&
[K^a,J^b]=\ri\varepsilon_{abc}K^c,\quad [K^a,K^b]=0,\\& [K^a,P^b]=2\ri(\delta^{ab}D-\varepsilon_{abc}J^c)\end{split}\end{gather}
where $\varepsilon_{abc}$ is the Levi-Civita symbol. In other words, they form a basis of the Lie algebra of conformal group C(3) defined in the 3$d$ Euclidean space.

We see that the required integrals of motion should belong to algebra c(3) whose basis elements are given by equation (\ref{QQ}). However,
 if all parameters $\lambda^a, \nu^a, \mu^a$ and $\omega$ are arbitrary, then the determining equations (\ref{de5}) and (\ref{de6}) are compatible iff $f({\bf x})=0, V({\bf x})=Const$, and equation (\ref{se}) is trivial. To obtain a non-trivial equation (\ref{se}) it is necessary to impose some constraints on these parameters and reduce the algebra c(3) to one of its subalgebras (we remind  that the first order integrals of motion have to form a Lie algebra \cite{Miller}).

 Thus to fix all non-equivalent versions of equations (\ref{de5}) and (\ref{de6})
it is necessary to find all non-equivalent sets of parameters $\lambda^a, \nu^a, \mu^a$ and $\omega$, which correspond to non-equivalent subalgebras of algebra c(3). The optimal system of such subalgebras can be defined up to the group of internal isomorphisms, which is group C(3).

Happily, the optimal system of subalgebras of c(3) are well known. To enumerate them we note that algebra c(3) is isomorphic to so(1,4), i.e., to the Lie algebra of the Poincar\'e group in 1+4 dimensional space. This isomorphism can be fixed by the following formulae
\begin{gather}M^{ab}=\varepsilon^{abc}J^c,\ M^{0a}=\frac12(K^a+P_a),\ M^{4a}=\frac12(K^a-P_a),\  M^{04}=D\label{iso}\end{gather}
where $M^{\mu\nu}$ with $\mu, \nu=0,1,2,3,4$ are basis elements of algebra so(1,4), satisfying the following relations:
\begin{gather}\label{13}[M^{\mu\nu},M^{\lambda\sigma}]=\ri (g^{\mu\sigma}M^{\nu\lambda}+g^{\nu\lambda}M^{\mu\sigma}-
g^{\mu\lambda}M^{\nu\sigma}-g^{\nu\sigma}M^{\mu\lambda})\end{gather}
were $g^{\mu\nu}=diag(1,-1,-1,-1,-1)$. One can make sure that commutation relations (\ref{13}) are consequences of (\ref{c(3)}), (\ref{iso}), and wise versa, relations (\ref{c(3)}) follows from (\ref{13}) and (\ref{iso}).

Optimal subalgebras of algebra so(1,4) have been classified  in paper \cite{Pate}. Using this classification and applying isomorphism (\ref{iso}) we obtain the following list of non-equivalent subalgebras:

One dimension subalgebras:
\begin{gather}\la{m1}\begin{split}&<M^{21}>,\quad  <M^{43}-M^{03}>, \quad
<M^{43}-M^{03}+M^{21}>,\\&<M^{43}+\alpha M^{21}>,\quad \ 0<\alpha^2\leq 1,\quad <M^{04}-\nu M^{21}>,\quad 0\leq\nu\leq1
  ;\end{split}\end{gather}
two dimension subalgebras: \begin{gather}\la{m2}\begin{split}&<M^{43}, M^{21}>,\ \  <M^{42}-M^{02}, M^{41}-M^{01}>,\ \ <M^{04}, M^{21}>,\\&  <M^{21}, M^{43}-M^{03}>,\ \ < M^{43}-M^{03},\ \ M^{40}+\alpha M^{21}>;\end{split}\end{gather}
three dimension subalgebras:
\begin{gather}\la{m3}\begin{split}&<M^{43}-M^{12}, M^{42}-M^{31}, M^{41}-M^{23}>, \\& < M^{12}\cos c-M^{04}\sin c, M^{42}-M^{02}, M^{41}-M^{01}>,\ \  <M^{12}, M^{23}, M^{31}>,\\& <M^{21}, M^{42}-M^{02}, M^{41}-M^{01}>,\ \ <M^{03}, M^{32}-M^{02}, M^{31}-M^{01}>,\\& <M^{04},M^{03},M^{43}>,\ \  <M^{43}+M^{03}, M^{42}+M^{02},  M^{41}+M^{01}>,\\&<M^{43}+M^{03}+M^{21}, M^{42}+M^{02}, M^{41}+M^{01}>,\ \ <M^{04}, M^{12}, M^{43}-M^{03}>;\end{split}\end{gather}
four dimension subalgebras:\begin{gather}\la{m4}\begin{split}&<M^{43}, M^{21}, M^{42}-M^{31}, M^{41}+M^{32}>,\ \ <M^{04}, M^{12}, M^{31}, M^{23}> ,\\&<M^{12}, M^{04}, M^{42}-M^{02}, M^{41}-M^{01}>,\ \ <M^{43}, M^{01}, M^{02}, M^{12}>\\& <M^{04}, M^{43}-M^{03}, M^{42}-M^{02}, M^{41}-M^{01}>,\ \ \\&<M^{12}, M^{43}-M^{03}, M^{42}-M^{02}, M^{41}-M^{01}>;\end{split}\end{gather}
five dimension subalgebra:\begin{gather}<M^{04}, M^{12},  M^{43}+M^{03}, M^{42}+M^{02}, M^{41}+M^{01}>;\la{m5}\end{gather}
six dimension subalgebras:\begin{gather}\la{m6}\begin{split}&<M^{12}, M^{31}, M^{23}, M^{43}-M^{03}, M^{42}-M^{02}, M^{41}-M^{01}>,\\& <M^{31}, M^{23}, M^{12}, M^{01}, M^{02}, M^{03}>, <M^{41}, M^{42}, M^{43}, M^{12}, M^{31}, M^{23}>;\end{split}\end{gather}
seven dimension subalgebra:
\begin{gather} <M^{40}, M^{12}, M^{31}, M^{23}, M^{43}-M^{03}, M^{42}-M^{02}, M^{41}-M^{01}>;\la{m7}\end{gather}
and ten dimension subalgebra:\begin{gather}<M^{41}, M^{42}, M^{43}, M^{01}, M^{02}, M^{03}, M^{12}, M^{31}, M^{23}, M^{04}>\la{m10}\end{gather}
where $M^{\mu\nu}$ with $\nu=0,1,...,4$ are operators defined by equations (\ref{iso}).

Formulae given above repeat (but also  slightly correct) the classification results presented in paper \cite{Pate}. Namely:
 \begin{itemize}
   \item The one dimension algebra spanned on $M^{43}-\alpha M^{01}=A+\alpha K$ presented in \cite{Pate},  is equivalent to another one dimensional algebra, i.e., $<M^{21}\cos c-M^{03}\sin c>$ presented there. We change it by the correct representative $<M^{43}+\alpha M^{21}>$, see (\ref{m1}).
       \item In contrast with \cite{Pate} we do not fix particular values $\alpha=0, 1$ and $c=\frac{\pi}2$ in (\ref{m1}) since they do not correspond to special integrals of motion.
       \item To simplify the form of functions presented in the following  Table 2 we change some algebras presented in \cite{Pate} by another but equivalent ones. In particular, the one dimensional algebra  $<M^{32}-M^{03}>$ is replaced by the equivalent algebra $<M^{43}-M^{03}>$.
  \end{itemize}

Any of the enumerated subalgebras corresponds to a  system of the determining equations (\ref{de5}) and (\ref{de6}). More exactly, any of the one dimensional algebras generates a system of such equations, the two dimension algebras generate pairs of such systems, and so on. And our classification  problem  is decoupled to 34 subproblems corresponding to subalgebras (\ref{m1})--(\ref{m10}).
 The related functions $\xi^i$ and parameters $\lambda^a, \omega$ are easily recovered using definitions (\ref{sol1}), (\ref{Q1}) and (\ref{iso}). They are presented in Table 1.

   \begin{center} Table 1. Functions $\xi^a$ and $\eta$ corresponding to basis elements of algebra so(1,4).

  \vspace{1mm}

    \begin{tabular}{ccccccc}
 \hline
 No&Operators&$\xi^1$ &$\xi^2$ &$\xi^3$& $\eta$&Non-zero parameters in (\ref{Q1}) \\
 \hline
 1&$ \ri M_{43}$&$\ x_1x_3$\ &$\ x_2x_3\ $&$\ \frac{s_3+1}
 {2}\ $&$\ \frac{3 x_3}2\ $&$\ \lambda_3=-\nu_3=\frac{1}2\ $\\\
2&$\ri M_{42}$&$x_1x_2$&$\frac{s_2+1}{2}$&$x_2x_3$&$\frac{3 x_2}2$&
$\lambda_2=-\nu_2=\frac{1}2$\\
3&$\ri M_{41}$&$\frac{s_1+1}{2}$&$x_1x_2$&$x_1x_3$&$\frac{3 x_1}2$&$\lambda_1=-\nu_1=
\frac{1}2$\\
4&$\ri M_{40}$&$x_1$&$x_2$&$x_3$&$0$&$\omega=1$\\
5&$\ri M_{32}$&0&$x_3$&$-x_2$&$0$&$\mu_1=-1 $\\
6&$\ri M_{31}$&$x_3$&0&$-x_1$&$0$&$\mu_2=1 $\\
7&$\ri M_{21}$&$x_2$&$-x_1$&0&$0$&$\mu_3=-1 $\\
8&$\ri M_{03}$&$x_1x_3$&$x_2x_3$&$\frac{s_3-1}{2}$&$\frac{3x_3}2$&
$\lambda_3=\nu_3=\frac{1}2$\\
9&$\ri M_{02}$&$x_1x_2$&$\frac{s_2-1}{2}$&$x_2x_3$&$\frac{3x_2}2$
&$\lambda_2=\nu_2=\frac{1}2$\\
10&$\ri M_{01}$&$\frac{s_1-1}{2}$&$x_1x_2$&$x_1x_3$&$\frac{3x_1}2$
&$\lambda_1=\nu_1=\frac{1}2$\\
\hline\hline
\end{tabular}
\end{center}
\vspace{1mm}
where $s_a=2x_a^2-r^2,\ a=1,2,3 $. Functions $\xi^i$ corresponding to basis elements of subalgebras enumerated above are evident linear combinations of ones given in the table.

\section{Solution of determining equations}
Thus to find all non-equivalent first order integrals of motion for
equation (\ref{se}) it is necessary and sufficient to solve
determining equations (\ref{de5}) and (\ref{de6}) where $\xi^a$ are
polynomials specified in (\ref{sol1}). Moreover, it is sufficient to
go over the reduced versions of $\xi^a$ which correspond to the
subalgebras of algebra c(3) enumerated in (\ref{m1})--(\ref{m10}).
The explicit forms of these functions corresponding to basis
elements (\ref{iso}), (\ref{QQ})  are presented  in Table 1.

Let us start with the one dimension subalgebra spanned on $M^{03}$
(we set $c=\frac{\pi}2$ in (\ref{m1})). The corresponding functions
$\xi^a$, $\eta$ and nonzero parameter $\lambda_3$  are presented in
the eighth  line of Table 1. Substituting these functions into
equation (\ref{de5}) we obtain the following equation for $f$:
\begin{gather}\la{de7}
2x_3(x_1f_1+x_2f_2+x_3f_3)-({r^2+1})f_3=4x_3f
\end{gather}
whose general solution has the following form:
\begin{gather}\label{f1}f({\bf x})={\tilde r^2}F\left(\frac{x_2}{x_1},
\frac{r^2-1}{\tilde r}\right).\end{gather}
Here $\tilde r^2=x_1^2+x_2^2$, and $F(.,.)$ is an arbitrary function of its arguments.

The next step is finding the corresponding potential $V({\bf x})$ which solves the related equation (\ref{de6}), i.e.,
\begin{gather}\label{V1}2x_3(x_1V_1+x_2V_2+x_3V_3)-
{(r^2+1)}V_3=3f_3\end{gather}
where $f$ is the function presented in (\ref{f1}). This linear inhomogeneous equation is  solved by the following function:
\begin{gather}\label{V2}V({\bf x})=3\tilde rD_2F+\tilde{F}\left(\frac{x_2}{x_1},
\frac{r^2-1}{\tilde r}\right)\end{gather}
were $D_2F$ is the derivative of function $F$ with respect to its second argument $\frac{r^2-1}{\tilde r}$ and $\tilde F(.,.)$ is one more arbitrary function of $\frac{x_2}{x_1}$ and $\frac{r^2-1}{\tilde r}$.

Thus Hamiltonian (\ref{H}) admits integral of motion $M_{03}$ iff function $f$ and potential $V$ are given by equations (\ref{f1}) and (\ref{V2}). We see that there exist a rather extended class of Hamiltonians (\ref{H}) admitting   $M_{03}$. The corresponding free elements $f$ and $V$ depend on two arbitrary functions and are presented by equations (\ref{f1}) and (\ref{V2}).

Let us specify these functions by requiring that the corresponding Hamiltonians admit the additional integral of motion $M_{21}$. In this case functions  (\ref{f1}) and (\ref{V2}) have to satisfy the following additional equations:
\begin{gather*}\la{ee1}
x_2f_1-x_1f_2=0,\quad  x_2V_1-x_1V_2=0,
\end{gather*}
see (\ref{de5}), (\ref{de6}) and line 7 of Table 1. In other words, functions (\ref{f1}) and (\ref{V2}) have to be independent on their first argument $\frac{x_2}{x_1}$. Thus functions $f$ and $V$ are reduced to the following forms:
\begin{gather}\label{ff}f({\bf x})=\tilde r^2 F\left(\frac{r^2-1}{\tilde r}\right), \quad V({\bf x})=3\tilde rF'+\tilde{F}\left(\frac{r^2-1}{\tilde r}\right).\end{gather}
The symbol
 $F'$  in (\ref{ff}) denotes the derivation of function $F\left(\frac{r^2-1}{\tilde r}\right)$ with respect to its argument  $\frac{r^2-1}{\tilde r}$.

If Hamiltonian (\ref{H}) admits two another integrals of motion, namely, $<M^{03}, M^{32}-M^{02}>$, then functions  $f({\bf x})$ (\ref{f1}) and $V{\bf x})$ (\ref{V2}) should satisfy the following additional conditions:
\begin{gather}\la{de12}\begin{split}&
2x_2(x_1f_1+x_2f_2+(x_3-1)f_3)-({r^2-1-2x_3})f_2=4x_2f,\\&
2x_2(x_1V_1+x_2V_2+(x_3-1)V_3)-({r^2-1-2x_3})V_2=3f_2.\end{split}
\end{gather}
Functions  (\ref{f1}) and  (\ref{V2}) solve equations (\ref{de12}) iff they are reduced to the following special form:
\begin{gather}\label{de13}f(x)=x_1^2F\left(\frac{r^2-1}{x_1}\right), \quad V=3x_1F'+\tilde{F}\left(\frac{r^2-1}{x_1}\right).\end{gather}

It appears that Hamiltonian (\ref{H}) with $f$ and $V$ given in (\ref{de13}) in fact admits three integrals of motion since it commutes  with both operators $M^{02}$ and $M^{32}$. This result is represented in Item 9 of Table 2 where we change index 3 to 1.

Thus we have gone over all optimal subalgebras of dimension one and two, which include basis element $M^{03}$ and are equivalent to (\ref{m1}) and (\ref{m2}) . There is only one three dimension subalgebra including $M^{03}$ in the list (\ref{m3}), namely, $<M^{03}, M^{32}-M^{02}, M^{31}-M^{01}>$
 To specify Hamiltonians (\ref{H}) which admit  this algebra  we look for  functions (\ref{de13}) satisfying the additional constraint
\begin{gather}\la{de14}\begin{split}&
2x_1(x_1f_1+x_2f_2+(x_3-1)f_3)-({r^2-1-2x_3})f_1=4x_1f,\\&
2x_1(x_1V_1+x_2V_2+(x_3-1)V_3)-({r^2-1-2x_3})V_1=3f_1\end{split}
\end{gather}
which are equations (\ref{de5}) and (\ref{de6})  corresponding to the difference of data from lines  6 and 10 of Table 1.

Substituting (\ref{de13}) into (\ref{de14}) and integrating the resultant equations we obtain:
\begin{gather}\label{fV1}f(x)=\mu({r^2-1})^2, \quad V=6\mu r^2+\nu\end{gather}
where $\mu$ and $\nu$ are arbitrary constants. These solutions are compatible with all other extensions of algebras $<M^{03}, M^{21}>$ and  $<M^{03}, M^{32}-M^{02}>$ enumerated  in (\ref{m4})--(\ref{m10}).

In analogous way we calculate integrals of motion forming the other algebras enumerated in (\ref{m1})--(\ref{m10}). The final classification results are presented in Table 2.

\begin{center}
Table 2. Functions $f$ and $V$ in Hamiltonians  (\ref{H}) and the corresponding integrals of motion.

\vspace{2mm}

\begin{tabular}{cccc}
 \hline
  No& f & V & Integrals of motion\\
  \hline
     1&$F(\tilde r^2,x_{3})$&$\tilde F(\tilde r^2,x_{3})$&$M^{21}$ \\

  2&$F(x_1,x_2) $&$\tilde F(x_1,x_3)$&$M^{43}-M^{03}$\\

3&$ r^2 F(\theta,r^\nu e^\varphi)$
 &
 ${\tilde F}(\theta,r^\nu e^\varphi)$& $M^{04}-\nu M^{21}$ \\

 4&$\tilde r^2F(\frac{r^2+1}{\tilde r},\omega)$
 &$\begin{array}{c}3\tilde r(D_1-\frac{\alpha x_3\tilde r}{2k_{+}k_-}D_2)F\\+
 \tilde{F}(\frac{r^2+1}{\tilde r},\omega)\end{array}$& $M^{43}+\alpha M^{21}$ \\
 5&$F({\tilde r}^2,x_{3}-\varphi)$ & $\tilde F({\tilde r}^2,x_{3}-\varphi )$&$M^{43}-M^{03}+M^{21}$\\

  6&$\tilde r^2 F(\frac{r^2+1}{\tilde r})$ &$ 3\tilde rF'+\tilde{F}(\frac{r^2+1}{\tilde r})$&$M^{43},\ M^{21}$  \\

   7& $F(\tilde r)$ &$\tilde F(\tilde r)$&$M^{21},M^{43}-M^{03}$ \\

   8&$\tilde r^2F( \tilde r^\alpha e^\varphi)$&$\tilde{F}(  \tilde r^\alpha e^\varphi)$&$M^{43}-M^{03},M^{40}+\alpha M^{21}$\\

 9&$r^2 F(\theta)$&$\tilde F(\theta)$&$ M^{40}, M^{12}$\\

10&$r^2F(\varphi)$ &$\tilde{F}(\varphi)  $
& $M^{04},M^{03},M^{43} $ \\

 11&$F(x_3)$&$\tilde F(x_3)$&$M^{41}-M^{01}, M^{42}-M^{02}, M^{12}$\\

  12&$F(r^2)$&$\tilde{F}(r^2)$&$M^{32},M^{31},M^{21} $\\

  13&$\mu r^2$&$\nu$&$M^{04},M^{12},M^{31},M^{23}$\\

  14&$\mu \tilde r^2$&$\nu$&$M^{40}, M^{21}, M^{43}, M^{03}$\\

         15&$\mu(r^2+1)^2
         $&$6\mu r^2+\nu$&$M^{41}, M^{42}, M^{43}, M^{21}, M^{31}, M^{32}$ \\

 16&$\mu(r^2-1)^2$&$6\mu r^2+\nu$&$ M^{01}, M^{02}, M^{03}, M^{21}, M^{31}, M^{32}$\\

 17&$\mu r^4$&$6\mu r^2+\nu$&$\begin{array}{c}M^{31}, M^{21}, M^{32}, M^{43}+M^{03}, \\ M^{42}+M^{02}, M^{41}+M^{01}\end{array}$\\
 \hline\hline
 \end{tabular}
 \end{center}

 \vspace{2mm}
  Here $\mu$, $\nu$ and $\alpha\neq0$ are arbitrary constants, $\varphi$ and $\theta$ are Euler angles,
\begin{gather*}
\begin{split}&
 \omega=
\alpha\arctan\frac{r^2+1}{2x_3}- \varphi,\quad k_\pm=\tilde r^2+(x_3\pm1)^2,
\end{split}
\end{gather*}
while $D_1 F(...), D_2 F(...)$ and $D F(...)$ denote the derivations of functions $F(...)$ with respect to their the first, second and the only element correspondingly.

Let us note that solutions presented in Item 17 of Table 2 correspond to Hamiltonian (\ref{H}) which can be reduced to the Hamiltonian with the constant mass and constant potential. It can be done using the inversion transformation  ${\bf x}\to \frac{\bf x}{r^2}$.

\section{Exact solutions for maximally superintegrable systems }
It was conjectured in \cite{w1}   that all maximally superintegrable systems with two degrees of freedom are exactly solvable, and till now there are no counterexamples for this conjecture. Let us note that for 3d systems the connections between superintegrability and exact solvability is much more complicated.

In this section we show that the maximally superintegrable PDM systems admitting first order integrals of motion are exactly solvable too, and find the corresponding exact solutions explicitly.

\subsection{System invariant w.r.t. algebra so(4)}
Consider Hamiltonian (\ref{H}) with functions $f$ and $V$ presented in  line 15 of Table 2:
\begin{gather}\la{H2}H=\mu\left(p_a(1+r^2)^2p_a-6 r^2+\nu\right).\end{gather}
The eigenvalue problem for this Hamiltonian can be written in the following form:
\begin{gather}\la{ep}\hat H\psi\equiv-\left(\p_a(1+r^2)^2\p_a+6 r^2\right)\psi=\tilde E\psi\end{gather} where
\begin{gather}\hat H=\frac1\mu(H-\nu)\ \  \text{and}\ \ \tilde E=\frac{E}\mu-\frac\nu\mu.\la{scal}\end{gather}

Equation (\ref{ep}) admits six integrals of motion $M^{AB}, \ A, B=1,2,3,4$:
\begin{gather}\la{im}\begin{split}&M^{ab}=x^ap^b-x^bp^a,\\
&M^{4a}=\frac12(r^2-1)p^a -x^ax^bp^b+\frac{3\ri}2x^a\end{split}\end{gather}
which satisfy the following commutation relations:
\begin{gather}\label{cr1}[M^{AB},M^{CD}]=\ri (\delta^{AC}M^{BD}+\delta^{BD}M^{AC}-
\delta^{AD}M^{BC}-\delta^{BC}M^{AD})\end{gather} where $\delta^{AB}$ is the Kronecker delta.
In other words, operators (\ref{im}) form a basis of algebra so(4).
Defining new basis
\[q_a=\frac12\left(M^{4a}+\frac12\varepsilon^{abc}M^{bc}\right),\quad
g_a=\frac12\left(-M^{4a}+\frac12\varepsilon^{abc}M^{bc}\right)\]  it is possible to decouple algebra o(4) to the direct sum of two algebras $so(3)$ since $q^a$ and $g^a$ satisfy
\[[q^a,q^b]=\ri\varepsilon^{abc}q^c,\quad [g^a,g^b]=\ri\varepsilon^{abc}g^c,\quad [q^a,g^b]=0.\]

Thus system (\ref{ep}), like the Hydrogen atom, admits six integrals of motion belonging to algebra so(4) and is maximally superintegrable. However, in contrast with the Fock symmetry of the Hydrogen atom, integrals of motion (\ref{im}) are  first order differential operators.

Using the mentioned symmetry it is possible to find eigenvalues $\tilde E$ algebraically, before solving equation (\ref{ep}). To do it we calculate Casimir operators of algebra o(4) for representation (\ref{im}):
\begin{gather}\la{CO}C_1=\frac12M^{AB}M^{AB}=2({\bf q}^2+{\bf g}^2)\equiv\frac14 (\hat H-9), \\C_2=\frac12\varepsilon_{abc}M^{4a}M^{bc}=2({\bf q}^2-{\bf g}^2)\equiv0.\end{gather}
where ${\bf q}^2=(q^1)^2+(q^2)^2+(q^3)^2$, etc.

In accordance with (\ref{CO}) the Hamiltonian eigenvalues can be expressed via eigenvalues of $C_1$. Since $C_2$ is trivial, eigenvalues of $C_1$ can take the following values (compare with  \cite{N2}, equation (7))
\begin{gather}\la{ev1}C_1\psi=4q(q+1)\psi=(n^2-1)\psi,\quad n=1,2,...\end{gather}
and so eigenvalues of Hamiltonians $H$ (\ref{H2}) and $\tilde H$ are:
\begin{gather}\la{ev2}E=\mu(4 n^2+5)+\nu,\quad \text{and}\quad \tilde E=4 n^2+5.\end{gather}

 The next step is to find eigenvectors of Hamiltonian (\ref{H2}) corresponding to eigenvalues (\ref{ev2}). Using the rotation invariance of (\ref{ep}) it is possible to separate variables. Introducing spherical variables and expanding solutions via spherical functions
 \begin{gather}\la{rv}\psi=\frac1r\sum_{l,m}\phi_{lm}(r)Y^l_m\end{gather}
 we obtain the following equations for radial functions
 \begin{gather}\la{re}\left(-(r^2+1)^2\left(\frac{\p^2}{\p r^2}-\frac{l(l+1)}{r^2}\right)-4r(r^2+1)\frac{\p}{\p r}-2r^2\right)\varphi_{lm}=\left(4 n^2+1\right)\varphi_{lm},\end{gather} where $l=0,1,2,...$ is the spectral parameter labeling eigenvectors of the squared orbital momentum.
 The square integrable solutions of these equations are:
 \begin{gather}\la{soll}\varphi_{lm}=C_{lm}^n(r^2+1)^{-n-\frac12}r^{l+1}
 {\cal F}\left(\left[-n+l+1, -n+\frac12\right], \left[\frac32+l\right], -r^2\right)\end{gather}
 where ${\cal F}(\cdots)$ is the hypergeometric function and $C_{lm}^n$ are integration constants. Solutions (\ref{soll}) tend to zero at infinity provided the first argument of the wave function is a non-positive integer. This condition is in accordance with (\ref{ev2})  provided $l\leq n-1$.

 Thus the maximally superintegrable system (\ref{ep}) is exactly solvable. Its exact solutions and spectrum of Hamiltonian (\ref{H2}) are given by equations (\ref{soll}) and (\ref{ev2}). Let us note that the sign of eigenvalues $E$ coincides with the sign of coupling constant $\mu$.

 \subsection{System invariant w.r.t. algebra so(1,3)}
 The next Hamiltonian we consider corresponds to functions  $f$ and $V$
 presented in  line 16 of Table 2:
\begin{gather}\la{H21}H=\mu\left(p_a(1-r^2)^2p_a-6 r^2+\nu\right).\end{gather}

The related eigenvalue problem is based on the following equation:
\begin{gather}\la{ep1}\hat H\psi\equiv-\left(\p_a(1-r^2)^2\p_a+6 r^2\right)\psi=\tilde E\psi\end{gather} where notations (\ref{scal}) are used.

In accordance with Table 2, equation (\ref{ep1}) admits six integrals of motion $M^{\mu\nu}, \ \mu, \nu=0, 1,2,3$:
\begin{gather}\la{im1}\begin{split}&M^{ab}=x^ap^b-x^bp^a,\quad a,b=1,2,3\\
&M^{0a}=\frac12(r^2+1)p^a -x^ax^bp^b+\frac{3\ri}2x^a\end{split}\end{gather}
which satisfy the following commutation relations:
\begin{gather}\label{cr11}[M^{\mu\nu},M^{\lambda\sigma}]=\ri (g^{\mu\lambda}M^{\nu\sigma}+g^{\nu\sigma}M^{\mu\lambda}-
g^{\mu\sigma}M^{\nu\lambda}-\delta^{\nu\lambda}M^{\mu\sigma})\end{gather} where $g^{\mu\nu}=\text{diag}(-1,1,1,1)$.

Thus integrals of motion (\ref{im1}) form a basis of algebra so(1,3). Calculating the corresponding Casimir operators, we obtain:
\begin{gather}\la{CO2}C_1=\frac12M^{ab}M^{ab}-M^{0a}M^{0a}=\frac14 (\hat H+9), \\C_2=\frac12\varepsilon_{abc}M^{0a}M^{bc}=0.\end{gather}
Like in previous section, the first Casimir operator  is proportional to the (shifted) Hamiltonian.

Let us suppose that operators (\ref{im1}) generate a unitary representation (IR) of group SO(1,3). The admissible eigenvalues $c_1$ and $c_2$ of Casimir operators $C_1$ and $C_2$ (\ref{CO2}) are given by the following formulae \cite{naimark}, \cite{IM}:
\begin{gather} c_1=1-j_0^2-j_1^2, \quad c_2=2\ri j_0j_1\end{gather} where $j_0$ and $j_1$ are quantum numbers labeling irreducible representations. Since the second Casimir operator $C_2$ is trivial, we have $c_1=0$ and so $j_0=0$. In this case there are two possibilities \cite{naimark}: either $j_1$ is an arbitrary imaginary number, and the corresponding representation belongs to the principal series, or $j_1$ is real number satisfying $|j_1|\leq1 $. In the latter case we have a representation belonging to the subsidiary series of IRs. In other words,
\begin{gather}\la{j1}j_1=i\lambda, \quad c_1=1-j_1^2=\lambda^2+1 \end{gather}
where $\lambda$ is an arbitrary real number, or, alternatively,
\begin{gather}\la{j2}0\leq j_1\leq 1, \quad c_1=1-j_1^2\end{gather}

In accordance with (\ref{CO2}) the related eigenvalues $\tilde E$ in (\ref{ep1}) can be expressed as:
\begin{gather}\tilde E=-5-j_1^2.\la{EE}\end{gather}
They correspond to the continuous spectrum, and so there are no
bound states. Moreover, the possible values of  $\tilde E$ should
belong to the following intervals:
\begin{gather}\la{E1}-6\leq\tilde E\leq-5 \qquad \text{(principal series)}\\\la{E2}-5\leq\tilde E<\infty \qquad \text{(subsidiary series)}\end{gather}

Thus we can find admissible energies applying purely group-theoretical arguments. Our conclusions can be supported by direct solution of equation (\ref{ep1}). Taking into account
the rotational invariance of equation (\ref{ep1}) it is convenient to expand its solutions via spherical harmonics, i.e., represent them in form (\ref{rv}). As a result we obtain the following radial equations  \begin{gather}\la{re1}\left(-(r^2-1)^2\left(\frac{\p^2}{\p r^2}-\frac{l(l+1)}{r^2}\right)-4r(r^2-1)\frac{\p}{\p r}-2r^2\right)\varphi_{lm}=(\tilde E+4)\varphi_{lm}.\end{gather}

There is an  exceptional point $r=1$ for equation (\ref{re1}) thus we can expect that its solution will have the corresponding singularity. Indeed, the general  solution of (\ref{re1}) looks as follows:
 \begin{gather}\la{soll1}\begin{split}&\varphi_{lm}=C_{lm}^k(1-r^2)^{-
 \frac12-k}r^{l+1}{\cal F}\left(\left[-k+l+1, -k+\frac12\right],
  \left[\frac32+l\right], r^2\right)\\&+\tilde C_{lm}^k(1-r^2)^{-
 \frac12-k}r^{-l}{\cal F}\left(\left[-k-l, -k+\frac12\right],
 \left[\frac12-l\right], r^2\right)\end{split}
 \end{gather}
  where $k=\frac12\sqrt{-\tilde E-5}$, and is singular at $r=1$.

 However, setting in  (\ref{soll1})  $\tilde C_{lm}^k=0$  and $k= j_1$ with $j_1$ satisfying (\ref{E1})  or (\ref{E2}) we come to solutions which are normalizable in the following metrics:
   \begin{gather}\la{mtrx}(\psi_1,\psi_2)=
   \int_0^1\psi_1^\dag\psi_2(r^2-1)^3dr.\end{gather}
   Moreover, the  expression under integral is equal to zero at $r=0$
   and $r=1$ provided $0\leq j_1^2<1$ or $j_1^2\leq0$.

    \subsection{Scale invariant system}
 The last system we consider corresponds to Line 14 of Table 2 were we set $\nu=0$, and so it is specified by the following Hamiltonian:
 \begin{gather} \la{last}H=p_a\tilde r^2p_a=-\tilde r^2\Delta-2x_\alpha\nabla_\alpha,\quad \alpha=1,2.\end{gather}
 This Hamiltonian is transparently invariant with respect to the simultaneous scaling of all independent variables, rotations in the plane 1-2 and shifts of $x_3$. Considering the eigenvalue problem for (\ref{last}) it is convenient to use the cylindrical variables $\tilde r=\sqrt {x_1^2+x_2^2}, \ \varphi=\arctan\frac{x_2}{x_1},\ x_3=z$ and expand solutions via eigenfunctions of $M^{12}$ and $P_3=-\ri\frac{\p}{\p z}$:
 \begin{gather*} \Psi=\exp[\ri(\kappa\varphi +\omega z)]\Phi_{\kappa\omega}(\tilde r),\quad \kappa=0,\pm1, \pm2,..., -\infty<\omega<\infty.\end{gather*}
 As a result we come to the following equations for radial functions $\Phi=\Phi_{\kappa\omega}(\tilde r)$:
 \begin{gather*}-\tilde r^2\left(\frac{\p^2\Phi}{\p \tilde r^2}+\omega^2\Phi\right)-3\tilde r\frac{\p \Phi}{\p \tilde r}=(\tilde E-\kappa^2.)\Phi\end{gather*}
 Square integrable (with the weight $\tilde r$) solutions of this equation are:
 \begin{gather}\la{soso}\Phi_{\kappa\omega}=\frac1{\tilde r}J_\alpha(\omega \tilde r),\quad \alpha=\kappa^2+1-\tilde E \end{gather}
 where $J_\alpha(\omega \tilde r)$ is Bessel function of the first kind.
 Functions (\ref{soso}) are normalizable and disappear at $\tilde r=0$ provided $\alpha\leq0$. The rescaled energies  $\tilde E$ continuously take the values $\kappa^2\leq\tilde E\leq\infty$.

\section{Discussion}
Thus we find all non-equivalent PDM Schr\"odinger equations which admit at least one first order integral of motion. These equations are presented by formulae (\ref{se}) and (\ref{H}) where $f$ and $V$ are functions collected in Table 2. Equations corresponding to lines 1 -- 12 of the table are rather general and include one or two arbitrary functions.

Any of  systems whose potentials are presented in lines 1 -- 5 admits only one integral of motion. Lines 6 -- 9 present integrable equations which admit pairs of commuting integrals of motion. Equations corresponding to lines 10 -- 16 admit even more of them and are superintegrable. Moreover, lines 13 -- 16 represent maximally superintegrable systems any of which admits four algebraically independent integrals of motion in addition to Hamiltonians (this number is maximal for systems with three degrees of freedom). For completeness in line 17 we present the mass  and  potential which are equivalent to constant ones.

Let us note that Hamiltonians (\ref{H2}) and (\ref{H21}) are equivalent to particular cases of Hamiltonians considered in \cite{balt0}-\cite{balt3}, see, for instance, Example 8 in \cite{balt2} for $A=0$, or Hamiltonians $\hat H_{LB}$ with $\mu=0$ in \cite{balt0}. However, just (\ref{H2}) and (\ref{H21}), in contrast with the more general Hamiltonians considered in \cite{balt0}-\cite{balt3}, admit six first order integrals of motion forming a basis of algebras so(4) and so(1,3). On the other hand, the superintegrable PDM quantum  systems, discussed in \cite{balt}--\cite{balt4} (and other PDM systems) can be obtained in frames of a classification of second order integrals of motion. This work is in progress.

The equivalence group of our classification problem is the 3$d$ conformal group whose generators are defined by equation (\ref{QQ}). Using this group any of the presented system can be propagated to an  entire family of equations. We remind that this group includes the following transformations:
\begin{itemize}
\item
shifts \begin{gather}\la{et1}x_a\to x_a'= x_a +\nu_a,\quad p_a\to p_a, \quad \psi\to \psi'=\psi\end{gather} generated by $P_a$;
\item
rotations \begin{gather}\la{et2}x_a\to R_{ab}x_b,\quad p_a\to R^{-1}_{ab}x_b, \quad \psi\to \psi'=\psi \end{gather}
where $R_{ab}$ is an orthogonal matrix. This transformation is generated by $J_i$;
\item
dilatation  \begin{gather}\la{et3}x_a\to x_a'=e^\lambda x_a,\quad p_a\to p_a'=e^{-\lambda}p_a, \quad \psi\to \psi'=e^{-\frac{3\lambda}2}\psi\end{gather}
generated by $D$;
\item
conformal transformations \begin{gather}\la{et4}\begin{split}&x_a\to x'_a=\frac{x_a-\mu_ar^2}{1-2\mu_a x_a+\mu^2r^2},\\& p_a\to p_a'=p_a-2\mu_aD+2\varepsilon_{abc}\mu_bJ_c+2\mu_a\mu_bK_b-\mu^2K_a, \\&\psi\to \psi'=\frac\psi{(1-2\mu_a x_a+\mu^2r^2)^\frac32}\end{split}\end{gather} where $\mu^2=\mu_1^2+\mu_2^2+\mu_3^2$. These transformations are generated by $K_a$.
\end{itemize}
Here $\psi$ is a solution of equation (\ref{se}), the other Greek letters denote  transformation parameters.

Transformations (\ref{et1})--(\ref{et4}) keep the generic form of equation (\ref{se}) and algebras of integrals of motion presented in the last column of Table 2. However, they change the explicit forms of functions $f$ and $V$ in Hamiltonian (\ref{H}) and the corresponding integrals of motion. In other words, any line of Table 2 presents a class of equivalent equations (\ref{se}) defined up to transformations (\ref{et1})--(\ref{et4}) and their products.

Thus we classify PDM Schr\"odinger equations which admit the first order integrals of motion. Some of these equations have unusual  symmetries being invariant w.r.t. the Lie algebra of  Lorentz group SO(1.3) (see section 5.2) and w.r.t. the Lie algebra of group SO(1,2) (see line 10 of Table 2). These "relativistic aspects" of the PDM Schr\"odinger equations are rather inspiring.

Like the Hydrogen atom, the system discussed in Section 5.1 is invariant w.r.t. algebra so(4). This property can be used to find the Hamiltonian spectra algebraically and to solve equation (\ref{ep}) exactly. Exact solutions of this and some other equations presented in Section 5.

Thus we  present the completed list of PDM equations admitting first order integrals of motion. Since such integrals of motion are generators of continuous symmetry groups, our results can be treated as a group classification of equations (\ref{se}) including two arbitrary elements, i.e., functions $f$ and V. The natural next step is to extend this result to the case of non-stationary equations including the time derivation. This work is in progress.

One more challenge for researchers is the classification of PDM
Hamiltonians admitting higher order integrals of motion. Preliminary
results in this field are presented in paper \cite{AG}.

\end{document}